\documentclass[smallextended]{svjour3}       

\RequirePackage{fix-cm}

\smartqed  

\usepackage{latexsym}
\usepackage{amsfonts}
\usepackage{amsmath}
\usepackage{amssymb}

\usepackage{epsfig}
\usepackage{enumitem}

\usepackage[linesnumbered,lined,ruled]{algorithm2e}
\usepackage{marginnote}
\usepackage{graphicx}
\usepackage{natbib}

\spnewtheorem{subcor}{}[corollary]{\bfseries}{\itshape}
\numberwithin{subcor}{corollary}

\spnewtheorem{subprop}{}[proposition]{\bfseries}{\itshape}
\numberwithin{subprop}{proposition}

\spnewtheorem{sublemma}{}[lemma]{\bfseries}{\itshape}
\numberwithin{sublemma}{lemma}

\spnewtheorem{subtheorem}{}[theorem]{\bfseries}{\itshape}
\numberwithin{subtheorem}{theorem}

\newcommand{\NN}{{\mathbb N}}
\newcommand{\RR}{{\mathbb R}}
\newcommand{\CC}{{\mathcal C}}
\newcommand{\II}{{\mathbb I}}

\usepackage{color}
\newcommand{\blue}{\textcolor{black}}

\usepackage{tikz} 
\usetikzlibrary{arrows.meta}
\usetikzlibrary{arrows}
\usetikzlibrary{patterns}
\usetikzlibrary{calc}
\usetikzlibrary{intersections}
\usetikzlibrary{decorations.fractals}
\usetikzlibrary{decorations.markings}
\usetikzlibrary{graphs}

\tikzset{
	vertex/.style={circle,fill=black,scale=0.6,label distance=0.15,font=\Large}
}

\journalname{}

\begin{document}
\title{Counting and optimising maximum phylogenetic diversity sets}

\author{Kerry Manson \and
		Charles Semple   \and
		Mike Steel
}


\institute{K. Manson \and C. Semple \and M. Steel 
	\at	Biomathematics Research Centre, School of Mathematics and Statistics, \\ University of Canterbury, Christchurch, New Zealand \\
    \email{kerry.manson@pg.canterbury.ac.nz} (corresponding author)
}


\maketitle

\begin{abstract}
In conservation biology, phylogenetic diversity (PD) provides a way to quantify the impact of the current rapid extinction of species  on the evolutionary `Tree of Life'.  This approach recognises that extinction not only removes species but also the branches of the tree on which unique features shared by the extinct species arose.   In this paper, we investigate three questions that are relevant to PD. The first asks how many sets of species of given size $k$ preserve the maximum possible amount of PD in a given tree.  The number of such maximum PD sets can be very large, even for moderate-sized phylogenies.  We provide a combinatorial characterisation of maximum PD sets, focusing on the setting where the branch lengths are ultrametric (e.g. proportional to time). This leads to a polynomial-time algorithm for calculating the number of maximum PD sets of size $k$ by applying a generating function; we also investigate the types of tree shapes that harbour the most (or fewest) maximum PD sets of size $k$. 
Our second question concerns optimising a linear function on the species (regarded as leaves of the phylogenetic tree) across all the maximum PD sets of a given size. Using the characterisation result from the first question, we show how this optimisation problem can be solved in polynomial time, even though the number of maximum PD sets can grow exponentially. 
Our third question considers a dual problem:  If $k$ species were to become extinct, then what is the largest possible {\em loss} of PD in the resulting tree? For this question, we describe a polynomial-time solution based on dynamical programming. 
\end{abstract}

\newpage

\section{Introduction}
\label{intro_sec}

Advances  in molecular genetics and computational techniques over recent decades have allowed biologists to reconstruct evolutionary relationships among thousands of species \citep{jetz14, up19}. However, as fast as this `Tree of Life' is being assembled, many of these species are heading to extinction because of anthropogenic impacts \citep{dav18}.   This extinction of species  also entails the loss of features and genetic variation through the differential pruning of the underlying \blue{tree structure}. The impact on this tree is often estimated by the reduced sum of edge lengths measured in evolutionary time \citep{fai92}.  For example, if all 575 bird species classified as `imperilled' were to disappear from the bird phylogeny (from $\sim$10,000 species), this would result in the loss of 2.7 billion years of evolution \citep{jetz14}.

\begin{figure}[h]
	\centering\begin{tikzpicture}[scale=0.75]
	
	\begin{scope}[every node/.style=vertex]
	\node at (-3.5,5.5) () [draw=none,fill=none] {$T_1$};
	\node at (-1.25,6) (rho) [label=above:$\rho$]{};
	\node at (-4,0) (x1) [label=below:$x_1$]{};
	\node at (-3,0) (x2) [label=below:$x_2$]{};
	\node at (-2,0) (x3) [label=below:$x_3$]{};
	\node at (-1,0) (x4) [label=below:$x_4$]{};
	\node at (0,0) (x5) [label=below:$x_5$]{};
	\node at (1,0) (x6) [label=below:$x_6$]{};
	\node at (2,0) (x7) [label=below:$x_7$]{};
	\node at (-3,4) (c) {};
	\node at (0.5,4) (d) {};
	\node at (-3.5,2) (f) {};
	\node at (-0.5,2) (g) {};
	\node at (1.5,1) (h) {};
	\end{scope}
	
	\path[draw,thick,loosely dotted]
	(rho) edge[solid] node[midway,above,draw=none,fill=none] {2} (c)
	(rho) edge[solid] node[midway,above,draw=none,fill=none] {2} (d)
	(c) edge node[midway,left,draw=none,fill=none] {2} (f)
	(c) edge[solid] node[midway,right,draw=none,fill=none] {4} (x3)
	(d) edge[solid] node[midway,left,draw=none,fill=none] {2} (g)
	(d) edge[solid] node[midway,right,draw=none,fill=none] {3} (h)
	(f) edge node[midway,left,draw=none,fill=none] {2} (x1)
	(f) edge node[midway,right,draw=none,fill=none] {2} (x2)
	(g) edge[solid] node[midway,left,draw=none,fill=none] {2} (x4)
	(g) edge node[midway,right,draw=none,fill=none] {2} (x5)
	(h) edge node[midway,left,draw=none,fill=none] {1} (x6)
	(h) edge[solid] node[midway,right,draw=none,fill=none] {1} (x7)
	;
	\end{tikzpicture}
	\caption{The minimal subtree connecting species $x_3$, $x_4$ and $x_7$ has a PD score of 16.}
	\label{maxPDexample1}
\end{figure}
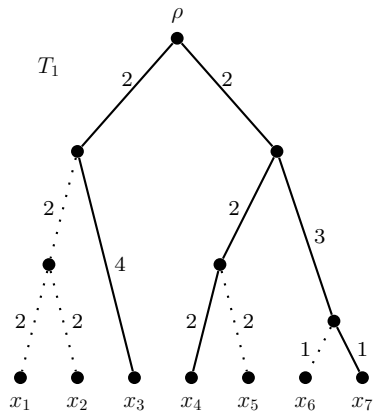

The ancestral relationships between a set of species are generally modelled using phylogenetic trees  \citep{fel04}, and one measure of how much of a tree is spanned by a subset of species is the phylogenetic diversity (PD) measure (precise definitions are provided in the next section;  here, we give an informal description). 
In simple terms, every non-empty set of species defines a minimal subtree which connects those species to the root of the tree, and the length of every branch in that subtree is summed to give a PD score for the set overall.
The greater the PD score, the more diverse a set of species is assumed to be. 
To illustrate, Fig.~\ref{maxPDexample1} shows the relative ancestry of the species $x_1, x_2, \dots, x_7$. 
Solid edges are those used in the calculation of the PD score for species $x_3$, $x_4$ and $x_7$.
Thus the PD score of $\{x_3,x_4,x_7\}$ is 16.
Note that the PD score of $\{x_4,x_7\}$ is 10.

An important concern of conservationists is preventing the extinction of species and the subsequent reduction of biodiversity.
For a phylogenetic tree, an extinction is represented by the removal of that species' leaf from the tree.
This also removes the edge which connected that species to the rest of the tree, lowering the PD score. 
In the case where more than one species becomes extinct, the combined effect can be much larger than the sum of individual extinctions.
We can interpret Fig.~\ref{maxPDexample1} as representing the extinction of $x_1, x_2, x_5$, and $x_6$.
Notice that the simultaneous extinction of $x_1$ and $x_2$ has caused the removal of a third edge, that connecting their least common ancestor to the rest of the tree.
These types of dependencies can lead to large differences in PD scores among sets of equal size.

Research has been conducted to assess the usefulness of the PD measure to inform conservation strategy.
To this end, sets of species which attain the maximum value of PD (for a given number of species) have been used as a benchmark against a measured response, to be contrasted with random selections of species \citep{tucker2019assessing}.  
However, the sets of a given size which maximise PD are not unique.
In applications of PD, we see the algorithms which generate such sets being run multiple times to account for this.
For example, \cite{molina2021maximum}[p. 586] and \cite{mazel2018prioritizing}[p. 7] both performed ten runs on each phylogenetic tree under consideration because:
\begin{quote} ``there are multiple subsets of size S that maximises PD in a phylogeny''
\end{quote}

\noindent and 

\begin{quote} ``For a given tree there are likely multiple, and possibly very many, sets of species with the same [maximum] PD'', 
\end{quote}
respectively.  Furthermore, \cite {mazel2017conserving}[p. 1021] noted that:
 \begin{quote} 
``this number will vary across simulations and could, in some case, be very large.''
\end{quote}

Although the non-uniqueness of these sets is known and accounted for, their total number is not well understood.
This leaves open questions about the most appropriate number of runs to perform in the above trials, and what the chances are that random selections of species also happen to form sets which maximise PD.
In this paper, we investigate mathematical questions concerning the enumeration of maximum PD sets of given size, as well as identifying the sets of species of given size whose extinction would result in  the largest loss of phylogenetic diversity.

\subsection{Outline of the paper}
We begin by stating in the next section the key mathematical definitions required in the paper. In Section~\ref{sec_maxPD}, we present a new characterisation of those sets which maximise PD for each possible size (Theorem~\ref{maxPDsufficientconditions}).
This characterisation allows us to count the number of such sets on any rooted phylogenetic tree, which previous methods could not achieve concisely.
Theorem~\ref{mTkpolynomialtime} sets out how this process may be achieved efficiently.
The conceptual approach from Theorem~\ref{maxPDsufficientconditions} is continued in Section~\ref{sec-linearfunction}, leading to Algorithm~\ref{alglinearfunction}, which selects, in polynomial time, one of these maximising sets that is optimal against a second measure.
In Section~\ref{sec-minPD}, we consider a dual problem: determining the greatest possible loss of PD if a certain number of species becomes extinct (this turns out to be equivalent to minimising PD for a given number of species).  A dynamic programming approach is used to solve this problem in polynomial time  for binary (or degree-bounded) rooted phylogenetic trees.

\section{Preliminaries}
\label{definitions}

\textbf{Phylogenetic trees.}
Let $X$ be a non-empty set of taxa (e.g. species), with $|X| = n$.
A \emph{rooted phylogenetic $X$-tree} is a rooted tree $T = (V,E)$, where $X$ is the set of leaves, and all edges are directed away from a distinguished root vertex $\rho$, and every non-leaf, \blue{non-root} vertex has out-degree at least 2.
In addition, when $|X| = 1$, the tree consisting of a single vertex is a rooted phylogenetic $X$-tree. 
All edges drawn in this paper will be directed down the page. 
If all non-leaf vertices of $T$ have out-degree 2, we say that $T$ is \emph{binary}.

Three types of restrictions on $T$ will be useful.
For $A \subseteq X$, the \emph{$A$-subtree} of $T$ is the minimal tree which connects the leaves of $A$ to the root vertex $\rho$.
\blue{In order for an $A$-subtree to be a phylogenetic tree, we suppress any non-root vertices with out-degree 1 which arise during its construction.}
For a set of vertices $V^\prime \subseteq V(T)$, the forest $T[V^\prime]$ is the restriction of $T$ to those vertices in $V^\prime$ and (directed) edges $(u^\prime,v^\prime) \in E(T)$, where $u^\prime, v^\prime \in V^\prime$.
A subtree of $T$ is \emph{pendant} if it can be disconnected from $\rho$ by deleting a single edge of $T$.
As shorthand, for an arbitrary set $A$ and element $x$, we write $A \cup x$ in place of $A \cup \{x\}$ and $A-x$ in place of $A \backslash \{x\}$.

For any vertex $v \in V(T)$, we write $x \in c_T(v)$ if $x \in X$ and the unique path from $\rho$ to $x$ includes $v$.  
That is, $c_T(v)$ is the set (\emph{cluster}) of leaves descended from $v$ in $T$. 
For the (directed) edge $e = (u,v)$, we define $c_T(e) = c_T(v)$.
If a cluster has size two or three, we call it, respectively, a \emph{cherry} or a \emph{triple}. 
A cluster of size four which contains two distinct cherries is called a \emph{fork}. 
In the rooted phylogenetic tree $T_1$ of Fig.~\ref{maxPDexample1} the set $\{x_1,x_2,x_3\}$ is a triple, the set $\{x_4,x_5,x_6,x_7\}$ is a fork, and each of $\{x_1,x_2\}$, $\{x_4,x_5\}$ and $\{x_6,x_7\}$ is a cherry.

\noindent\textbf{Phylogenetic diversity.}
The edges of every rooted phylogenetic tree considered in this paper are positively weighted.  
Let $T$ be a rooted phylogenetic $X$-tree, and let $\ell: E(T) \rightarrow \mathbb{R}^{>0}$ be a function which assigns a positive real-valued length $\ell(e)$ to each edge $e \in E(T)$.
Suppose that $u,v \in V(T)$ are two vertices of $T$ connected by a directed path from $u$ to $v$ (this path is unique if it exists).
Then the \blue{\emph{distance from $u$ to $v$}}, denoted  $d(u,v)$, is the sum of the lengths of the edges in this path. 
If an edge $e$ is subdivided into two edges $e_1$ and $e_2$, we require $\ell(e_1) + \ell(e_2) = \ell(e)$.
If $\ell$ is such that for any two distinct leaves $x$ and $y$ we have $d(\rho,x) = d(\rho,y)$, we say that $\ell$ satisfies the \emph{ultrametric} condition.

For a non-empty subset $Y$ of $X$, we define the \emph{phylogenetic diversity} of $Y$ on $T$, denoted by $PD_{(T,\ell)}(Y)$, to be the sum of the edge lengths of the \blue{$Y$-subtree}.
That is,
$$PD_{(T,\ell)}(Y) = \sum_{\substack {e \in E(T):\\ c_T(e)\cap Y\neq\emptyset}} \ell(e).$$
It will be usual for us to remove the subscript notation and write $PD(Y)$ when it is clear which rooted phylogenetic tree and edge length function we refer to.
We also write $PD(T)$ to denote the phylogenetic diversity of the entire $X$-tree $T$, in place of \blue{$PD_{(T,\ell)}(X)$}.

\begin{figure}[h]
	\centering
	\begin{tikzpicture}[scale=0.75]
	
	\begin{scope}[every node/.style=vertex]
	\node at (4.5,5.5) () [draw=none,fill=none] {$T_1$};
	\node at (6.75,6) (2rho) [label=above:$\rho$]{};
	\node at (4,0) (2x1) [label=below:$x_1$]{};
	\node at (5,0) (2x2) [label=below:$x_2$]{};
	\node at (6,0) (2x3) [label=below:$x_3$]{};
	\node at (7,0) (2x4) [label=below:$x_4$]{};
	\node at (8,0) (2x5) [label=below:$x_5$]{};
	\node at (9,0) (2x6) [label=below:$x_6$]{};
	\node at (10,0) (2x7) [label=below:$x_7$]{};
	\node at (5,4) (2c) {};
	\node at (8.5,4) (2d) {};
	\node at (4.5,2) (2f) {};
	\node at (7.5,2) (2g) {};
	\node at (9.5,1) (2h) {};
	\node at (12.5,5.5) () [draw=none,fill=none] {$T_1$};
	\node at (14.75,6) (3rho) [label=above:$\rho$]{};
	\node at (12,0) (3x1) [label=below:$x_1$]{};
	\node at (13,0) (3x2) [label=below:$x_2$]{};
	\node at (14,0) (3x3) [label=below:$x_3$]{};
	\node at (15,0) (3x4) [label=below:$x_4$]{};
	\node at (16,0) (3x5) [label=below:$x_5$]{};
	\node at (17,0) (3x6) [label=below:$x_6$]{};
	\node at (18,0) (3x7) [label=below:$x_7$]{};
	\node at (13,4) (3c) {};
	\node at (16.5,4) (3d) {};
	\node at (12.5,2) (3f) {};
	\node at (15.5,2) (3g) {};
	\node at (17.5,1) (3h) {};
	
	\end{scope}
	
	\path[draw,thick,loosely dotted]
	(2rho) edge[solid] node[midway,above,draw=none,fill=none] {2} (2c)
	(2rho) edge[solid] node[midway,above,draw=none,fill=none] {2} (2d)
	(2c) edge[solid] node[midway,left,draw=none,fill=none] {2} (2f)
	(2c) edge[solid] node[midway,right,draw=none,fill=none] {4} (2x3)
	(2d) edge node[midway,left,draw=none,fill=none] {2} (2g)
	(2d) edge[solid] node[midway,right,draw=none,fill=none] {3} (2h)
	(2f) edge node[midway,left,draw=none,fill=none] {2} (2x1)
	(2f) edge[solid] node[midway,right,draw=none,fill=none] {2} (2x2)
	(2g) edge node[midway,left,draw=none,fill=none] {2} (2x4)
	(2g) edge node[midway,right,draw=none,fill=none] {2} (2x5)
	(2h) edge[solid] node[midway,left,draw=none,fill=none] {1} (2x6)
	(2h) edge node[midway,right,draw=none,fill=none] {1} (2x7)
	(3rho) edge node[midway,above,draw=none,fill=none] {2} (3c)
	(3rho) edge[solid] node[midway,above,draw=none,fill=none] {2} (3d)
	(3c) edge node[midway,left,draw=none,fill=none] {2} (3f)
	(3c) edge node[midway,right,draw=none,fill=none] {4} (3x3)
	(3d) edge[solid] node[midway,left,draw=none,fill=none] {2} (3g)
	(3d) edge[solid] node[midway,right,draw=none,fill=none] {3} (3h)
	(3f) edge node[midway,left,draw=none,fill=none] {2} (3x1)
	(3f) edge node[midway,right,draw=none,fill=none] {2} (3x2)
	(3g) edge node[midway,left,draw=none,fill=none] {2} (3x4)
	(3g) edge[solid] node[midway,right,draw=none,fill=none] {2} (3x5)
	(3h) edge[solid] node[midway,left,draw=none,fill=none] {1} (3x6)
	(3h) edge[solid] node[midway,right,draw=none,fill=none] {1} (3x7)
	;
	\end{tikzpicture}
	\caption{A size-3 maxPD set $\{x_2,x_3,x_6\}$ and a size-3 minPD set $\{x_5,x_6,x_7\}$ for $T_1$. Solid lines indicate the $\{x_2,x_3,x_6\}$- and $\{x_5,x_6,x_7\}$-subtrees respectively. Hence $PD(\{x_2,x_3,x_6\}) = 16$, and $PD(\{x_5,x_6,x_7\}) = 11$.}
	\label{maxPDexample2}
\end{figure}
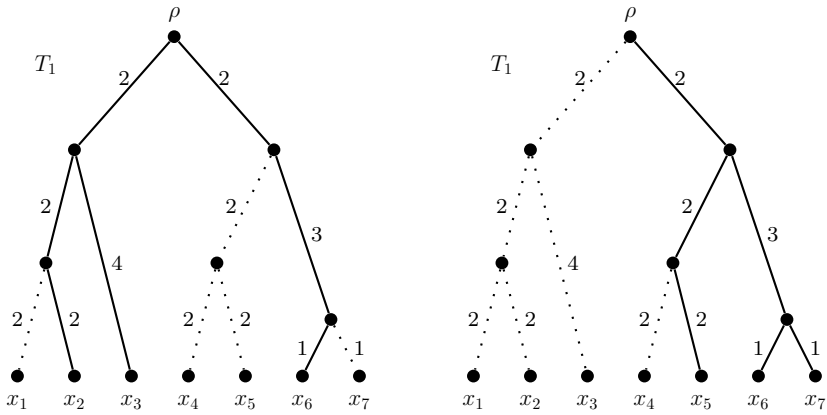

Let $T$ be a rooted phylogenetic $X$-tree whose edges are assigned a positive real-valued weighting, and let $A \subseteq X$.
If $|A|=k$ and $PD(A) \geq PD(Y)$ for all $Y \subseteq X$ with $|Y| = k$, then we call $A$ a \emph{size-$k$ maxPD set}.
Similarly, if $|A|=k$ and $PD(A) \leq PD(Y)$ for all $Y \subseteq X$ with $|Y| = k$, then we call $A$ a \emph{size-$k$ minPD set}.
To illustrate, Fig.~\ref{maxPDexample2} shows an example of a size-3 maxPD and an example of a size-3 minPD sets for the same rooted phylogenetic tree.

\section{The number of maxPD sets on rooted phylogenetic trees}
\label{sec_maxPD}

Given a rooted phylogenetic $X$-tree $T$, with $|X| =n$ and a weighting on $E(T)$, a natural question is to find a subset $Y$ of $X$ of size $t$ whose extinction minimises PD loss.
The solution to this question is to take $Y$ to be $X-W$, where $W$ is a subset of $X$ of size $n - t$ that maximises $PD(W)$.
It turns out that a greedy algorithm provably constructs such sets $W$ of $k = n - t$ leaves \citep{pardi2005species, steel2005phylogenetic}. 
This result relies on an underlying combinatorial `strong exchange property'  that induces a greedoid structure on maximal PD sets of given size.

Although a greedy algorithm will output a maxPD set, it does not give a clear indication of how many distinct maxPD sets exist for $T$.
Such an algorithm begins with an empty set of leaves and iteratively adds $k$ leaves, based on which leaf adds most to the running total of PD at each iteration.
There may be multiple steps at which a choice has to be made between equally-good options.
By altering the procedure for breaking ties when they occur, it is possible to discover numerous size-$k$ maxPD sets.
This effect is most pronounced for rooted phylogenetic $X$-trees satisfying the ultrametric condition.
For example, Fig.~\ref{maxPDexample1} and Fig.~\ref{maxPDexample2} show two of the 20 different size-3 maxPD sets for the rooted phylogenetic tree $T_1$.

All possible maxPD sets can be obtained by using a greedy algorithm, by taking each option separately when presented with ties \citep[Theorem 1]{steel2005phylogenetic}.
However, this process can become quite involved even for small phylogenetic trees. 
Moreover, each maxPD set could be counted multiple times, as greedy algorithms sometimes select the same set of leaves in different orders.

In this section, we present a more straightforward method for determining exactly how many maxPD sets exist on a given rooted phylogenetic $X$-tree whose edge lengths satisfy the ultrametric condition.
Firstly, by deleting certain edges near the root vertex, we partition the leaf set into disjoint subsets. Then we use a generating function which takes the sizes of these subsets and outputs the number of maxPD sets as a coefficient.

\subsection{Counting maxPD sets in an ultrametric context}\label{maxPDultrametric}

We restrict our attention to the problem of counting maxPD sets on a rooted phylogenetic $X$-tree $T$ whose edge lengths satisfy the ultrametric condition.
Suppose $T = (V,E)$, and let $1 \leq k \leq |X|$.
It turns out that the minimal subtrees of $T$ connecting size-$k$ maxPD sets to the root all contain particular subsets of edges. 
Furthermore, these common edges induce a subtree of $T$ containing the root vertex.
Our approach is to determine which edges of $T$ will be in common to all size-$k$ maxPD sets.
From there, we can enumerate these maxPD sets by analysing the forest that results from deleting the common edges. 

For example, all twenty size-3 maxPD sets of $T_1$ from Fig.~\ref{maxPDexample1} (with a score of 16) can be found by checking the 35 possible sets of 3 leaves.
Comparing these, we see that all of the minimal subtrees of $T_1$ that connect a size-3 maxPD set to the root of $T_1$ contain both edges incident with the root, as well as exactly 3 out of the 4 edges descending from the two highest non-root vertices.

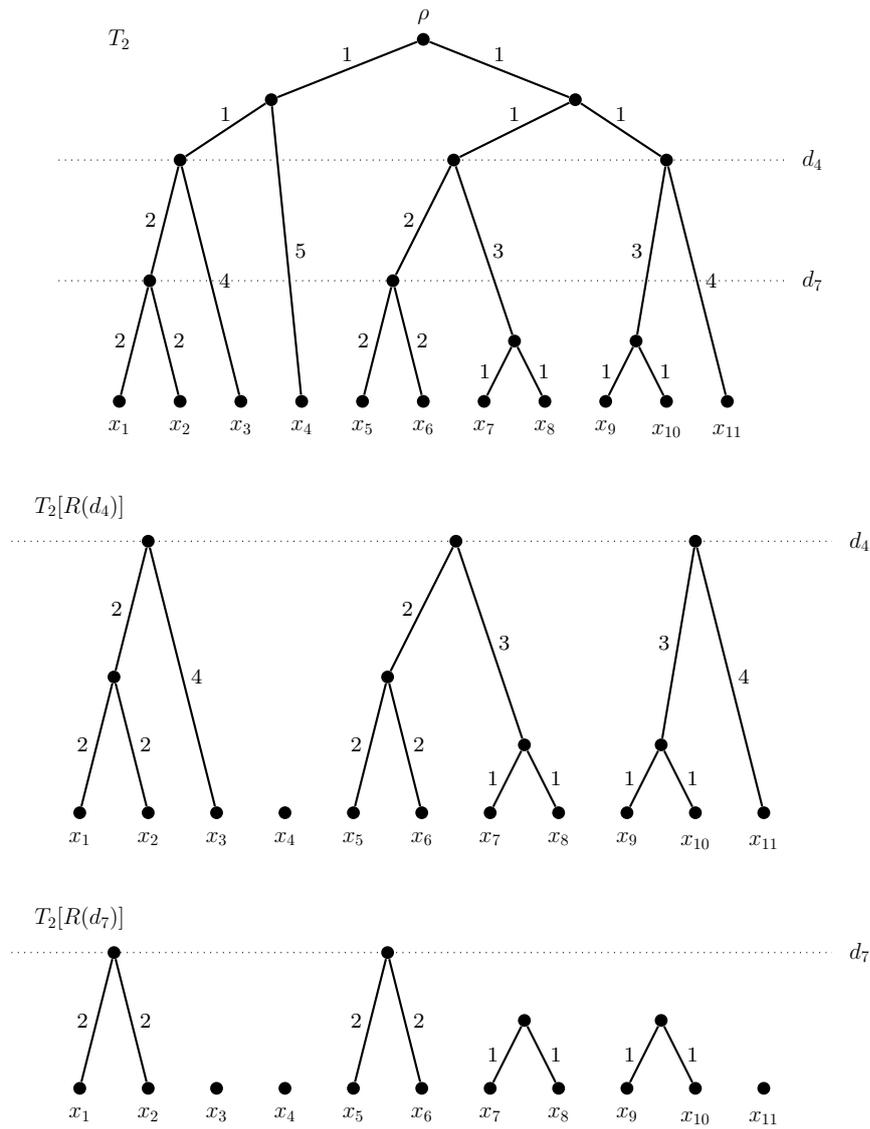
\begin{figure}[h]
	\centering
	
	\begin{tikzpicture}[scale=0.8]
	\begin{scope}[every node/.style=vertex]
	\node at (-5,6) [draw=none,fill=none] {$T_2$};
	\node at (6.4,2) [draw=none,fill=none] {$d_7$};
	\node at (6.4,4) [draw=none,fill=none] {$d_4$};
	\node at (-5,0) (x1) [label=below:$x_1$]{};
	\node at (-4,0) (x2) [label=below:$x_2$]{};
	\node at (-3,0) (x3) [label=below:$x_3$]{};
	\node at (-2,0) (x4) [label=below:$x_4$]{};
	\node at (-1,0) (x5) [label=below:$x_5$]{};
	\node at (0,0) (x6) [label=below:$x_6$]{};
	\node at (1,0) (x7) [label=below:$x_7$]{};
	\node at (2,0) (x8) [label=below:$x_8$]{};
	\node at (3,0) (x9) [label=below:$x_9$]{};
	\node at (4,0) (x10) [label=below:$x_{10}$]{};
	\node at (5,0) (x11) [label=below:$x_{11}$]{};
	\node at (-2.5,5) (a) {};
	\node at (2.5,5) (b) {};
	\node at (-4,4) (c) {};
	\node at (0.5,4) (d) {};
	\node at (4,4) (e) {};
	\node at (-4.5,2) (f) {};
	\node at (-0.5,2) (g) {};
	\node at (1.5,1) (h) {};
	\node at (3.5,1) (i) {};
	\node at (0,6) (rho) [label=above:$\rho$]{};
	
	\end{scope}
	\path[draw,thick]
	(rho) edge node[midway,above,draw=none,fill=none] {1} (a)
	(rho) edge node[midway,above,draw=none,fill=none] {1} (b)
	(a) edge node[midway,above,draw=none,fill=none] {1} (c)
	(a) edge node[midway,right,draw=none,fill=none] {5} (x4)
	(b) edge node[midway,above,draw=none,fill=none] {1} (d)
	(b) edge node[midway,above,draw=none,fill=none] {1} (e)
	(c) edge node[midway,left,draw=none,fill=none] {2} (f)
	(c) edge node[midway,right,draw=none,fill=none] {4} (x3)
	(d) edge node[midway,left,draw=none,fill=none] {2} (g)
	(d) edge node[midway,right,draw=none,fill=none] {3} (h)
	(e) edge node[midway,left,draw=none,fill=none] {3} (i)
	(e) edge node[midway,right,draw=none,fill=none] {4} (x11)
	(f) edge node[midway,left,draw=none,fill=none] {2} (x1)
	(f) edge node[midway,right,draw=none,fill=none] {2} (x2)
	(g) edge node[midway,left,draw=none,fill=none] {2} (x5)
	(g) edge node[midway,right,draw=none,fill=none] {2} (x6)
	(h) edge node[midway,left,draw=none,fill=none] {1} (x7)
	(h) edge node[midway,right,draw=none,fill=none] {1} (x8)
	(i) edge node[midway,left,draw=none,fill=none] {1} (x9)
	(i) edge node[midway,right,draw=none,fill=none] {1} (x10)
	;
	\draw [dotted] (-6,2) -- (6,2);
	\draw [dotted] (-6,4) -- (6,4);
	
	\end{tikzpicture}
	
	\begin{tikzpicture}[scale=0.9]
	\begin{scope}[every node/.style=vertex]
	\node at (-5,4.5) [draw=none,fill=none] {$T_2[R(d_4)]$};
	\node at (6.4,4) [draw=none,fill=none] {$d_4$};
	\node at (-5,0) (x1) [label=below:$x_1$]{};
	\node at (-4,0) (x2) [label=below:$x_2$]{};
	\node at (-3,0) (x3) [label=below:$x_3$]{};
	\node at (-2,0) (x4) [label=below:$x_4$]{};
	\node at (-1,0) (x5) [label=below:$x_5$]{};
	\node at (0,0) (x6) [label=below:$x_6$]{};
	\node at (1,0) (x7) [label=below:$x_7$]{};
	\node at (2,0) (x8) [label=below:$x_8$]{};
	\node at (3,0) (x9) [label=below:$x_9$]{};
	\node at (4,0) (x10) [label=below:$x_{10}$]{};
	\node at (5,0) (x11) [label=below:$x_{11}$]{};
	\node at (-4,4) (c) {};
	\node at (0.5,4) (d) {};
	\node at (4,4) (e) {};
	\node at (-4.5,2) (f) {};
	\node at (-0.5,2) (g) {};
	\node at (1.5,1) (h) {};
	\node at (3.5,1) (i) {};
	
	\end{scope}
	\path[draw,thick]
	(c) edge node[midway,left,draw=none,fill=none] {2} (f)
	(c) edge node[midway,right,draw=none,fill=none] {4} (x3)
	(d) edge node[midway,left,draw=none,fill=none] {2} (g)
	(d) edge node[midway,right,draw=none,fill=none] {3} (h)
	(e) edge node[midway,left,draw=none,fill=none] {3} (i)
	(e) edge node[midway,right,draw=none,fill=none] {4} (x11)
	(f) edge node[midway,left,draw=none,fill=none] {2} (x1)
	(f) edge node[midway,right,draw=none,fill=none] {2} (x2)
	(g) edge node[midway,left,draw=none,fill=none] {2} (x5)
	(g) edge node[midway,right,draw=none,fill=none] {2} (x6)
	(h) edge node[midway,left,draw=none,fill=none] {1} (x7)
	(h) edge node[midway,right,draw=none,fill=none] {1} (x8)
	(i) edge node[midway,left,draw=none,fill=none] {1} (x9)
	(i) edge node[midway,right,draw=none,fill=none] {1} (x10)
	;
	
	\draw [dotted] (-6,4) -- (6,4);
	
	\end{tikzpicture}
	
	\begin{tikzpicture}[scale=0.9]
	\begin{scope}[every node/.style=vertex]
	\node at (-5,2.5) [draw=none,fill=none] {$T_2[R(d_7)]$};
	\node at (6.4,2) [draw=none,fill=none] {$d_7$};
	\node at (-5,0) (x1) [label=below:$x_1$]{};
	\node at (-4,0) (x2) [label=below:$x_2$]{};
	\node at (-3,0) (x3) [label=below:$x_3$]{};
	\node at (-2,0) (x4) [label=below:$x_4$]{};
	\node at (-1,0) (x5) [label=below:$x_5$]{};
	\node at (0,0) (x6) [label=below:$x_6$]{};
	\node at (1,0) (x7) [label=below:$x_7$]{};
	\node at (2,0) (x8) [label=below:$x_8$]{};
	\node at (3,0) (x9) [label=below:$x_9$]{};
	\node at (4,0) (x10) [label=below:$x_{10}$]{};
	\node at (5,0) (x11) [label=below:$x_{11}$]{};
	\node at (-4.5,2) (f) {};
	\node at (-0.5,2) (g) {};
	\node at (1.5,1) (h) {};
	\node at (3.5,1) (i) {};
	
	\end{scope}
	\path[draw,thick]
	(f) edge node[midway,left,draw=none,fill=none] {2} (x1)
	(f) edge node[midway,right,draw=none,fill=none] {2} (x2)
	(g) edge node[midway,left,draw=none,fill=none] {2} (x5)
	(g) edge node[midway,right,draw=none,fill=none] {2} (x6)
	(h) edge node[midway,left,draw=none,fill=none] {1} (x7)
	(h) edge node[midway,right,draw=none,fill=none] {1} (x8)
	(i) edge node[midway,left,draw=none,fill=none] {1} (x9)
	(i) edge node[midway,right,draw=none,fill=none] {1} (x10)
	;
	\draw [dotted] (-6,2) -- (6,2);
	
	\end{tikzpicture}
	
	\caption{A rooted phylogenetic tree $T$, and the forests $T[R(d_4)]$ and $T[R(d_7)]$ corresponding to the branching values $4$ and $7$. The branching distances $d_4$ and $d_7$ are indicated by horizontal dotted lines.}
	\label{fig3}
\end{figure}

We extend the metaphor that the ultrametric condition produces clock-like trees and consider time to run down the page. 
Vertices at the same height are therefore contemporary and, in particular, the leaves are in the present.
Let $d$ be a non-negative real number and let $$R(d) = \{v \in V: d(v,x) \leq d \textrm{~for~some~} x \in X\}$$ be the set of \emph{recent} vertices of $T$ that are at most $d$ units of time from the present.
Let $c(d)$ be the number of connected components in $T[R(d)]$.
If there exists a distance $d$ such that $c(d) = k$, we define $d_k = \min \{d \in \RR : c(d) = k\}$.
Note that $d_k$ may not be defined for all $k < n$.
However, if $d_k$ is defined, we call $k$ a \emph{branching value}, and $d_k$ a \emph{branching distance}.
In other words, $d_k$ is the most recent time for which $T[R(d)]$ has exactly $k$ connected components, if such a time exists.
For example, the rooted phylogenetic tree $T_2$ in Fig.~\ref{fig3} has $\{1, 2, 4, 7, 9, 11\}$ as its set of branching values.
The forests $T_2[R(d_4)]$ and $T_2[R(d_7)]$ are shown below $T_2$ in the same figure.

If $k$ is not a branching value, we will be interested in the nearest integers which are. 
We write $k^+$ to denote the smallest branching value of at least $k$, and $k^-$ to denote the largest branching value of at most $k$.
Note that $k=1$ and $k = |X|$ are branching values, so that $k^+$ and $k^-$ are well-defined.
If $k$ is a branching value, then $k^- = k = k^+$.
Theorem~\ref{maxPDsufficientconditions} gives a characterisation of maxPD sets of a rooted phylogenetic tree $T$ in terms of the forests $T[R(d_{k^-})]$ and $T[R(d_{k^+})]$.

We first prove Lemma~\ref{PDupperbound}.
Let $X = \{x_1,\dots,x_n\}$. 
We define $T^\prime_k = (V^\prime,E^\prime)$ to be the rooted tree derived from $T$ (by adding vertices to subdivide edges as necessary) where, for each $x_i \in X$, there is a vertex $v_i$ on the path $\rho$ to $x_i$ for which $d(v_i,x_i) = d_{k^-}$.
Since $T^\prime_k$ is derived from $T$ solely by subdivision of edges, $PD_T(A) = PD_{T^\prime_k}(A)$.
Let $V^\prime_{Top}(k) = \{v \in V^\prime : d(\rho,v) \leq d_1 - d_{k^-}\}$, and let $\hat{T}_k$ be the \blue{rooted tree} $T^\prime_k[V^\prime_{Top}(k)]$.

\begin{lemma}\label{PDupperbound}
	Let $T$ be a rooted phylogenetic $X$-tree whose edge lengths satisfy the ultrametric condition, and let $A \subseteq X$ with $|A| = k$.
	Then 
	$$PD_T(A) \leq PD(\hat{T}_k) + kd_{k^-}.$$
\end{lemma}

\begin{proof}
	Let $A = \{x_1,\dots,x_k\}$ be a size-$k$ subset of $X$.
	Each element $x_i$ of $A$ contributes at most $d(\rho,x_i)$ to the total of $PD_{T^\prime_k}(A)$.
	We separate the path from $\rho$ to $x_i$ within $T^\prime_k$ into two parts at vertex $v_i$. 
	Hence $$d(\rho,x_i) = d(\rho,v_i) + d(v_i,x_i) =  d(\rho,v_i) + d_{k^-}.$$
	
	For all $1 \leq i \leq k$, the path from $\rho$ to $v_i$ lies within $\hat{T}_k$.
	Therefore the total contribution of these paths to $PD_{T^\prime_k}(A)$ cannot exceed $PD(\hat{T}_k)$. 
	This means $PD_{T^\prime_k}(A)$ must be less than or equal to $PD(\hat{T}_k)$ plus a contribution of (at most) $d_{k^-}$ from each of the $k$ elements of $A$.
	Thus  $PD_{T^\prime_k}(A) \leq PD(\hat{T}_k) + kd_{k^-}$, and the lemma holds.
\qed
\end{proof}

\blue{\begin{lemma}\label{mainlemma}
	Let $T$ be a rooted phylogenetic $X$-tree whose edge lengths satisfy the ultrametric condition.
	Let $A \subseteq X$ with $|A| = k$, and let $d$ be a branching distance of $T$.
	If one component of $T[R(d)]$ contains no members of $A$, while a second component of $T[R(d)]$ contains two or more distinct members of $A$, then $A$ is not a size-$k$ maxPD set.
\end{lemma}}

\begin{proof}
	\blue{Assume some component of $T[R(d)]$ contains two (distinct) leaves, say $x_1$, $x_2$, of $A$.
	The PD contribution of adding $x_1$ to $A - x_1$ cannot exceed $d$ because all edges of $T$ in the path from $\rho$ to $x_2$ have already been counted towards $PD(A - x_1)$.
	In particular, $PD(A) - PD(A - x_1) \leq d$. }
	 
\blue{	Now let $y$ be a leaf in a component of $T[R(d)]$ which contains no member of $A$.
	The shortest defined distance from a vertex in the $(A - x_1)$-subtree to $y$ must exceed $d$.
	(If not, there would be some vertex of the $(A - x_1)$-subtree in the same component of $T[R(d)]$ as $y$.)
	Hence $PD((A - x_1) \cup y) > PD(A)$. 
	Since $|(A - x_1) \cup y| = |A|$, the set $A$ is not a size-$k$ maxPD set.
	\qed}
	\end{proof}

\begin{theorem}\label{maxPDsufficientconditions}
	Let $T$ be a rooted phylogenetic $X$-tree whose edge lengths satisfy the ultrametric condition.
	Let $A \subseteq X$, and let $|A| = k$.
	Then $A$ is a size-$k$ maxPD set if and only if $A$ contains at least one leaf from each component of $T[R(d_{k^-})]$, and at most one leaf from each component of  $T[R(d_{k^+})]$. 
\end{theorem}

\begin{proof}
	First suppose that $A$ is a size-$k$ maxPD set.
	Assume some component of $T[R(d_{k^+})]$ contains two (distinct) leaves, say $x_1$,$x_2$, of $A$.
	Since $k^+ \geq k$, there must be some component of $T[R(d_{k^+})]$ which has no leaf in $A$. 
	\blue{Then, by Lemma \ref{mainlemma}, the set $A$ cannot be a maxPD set, contradicting our initial supposition.}
	Thus $A$ contains at most one leaf from each component of $T[R(d_{k^+})]$. 
	
	Next, assume that some component of $T[R(d_{k^-})]$ contains no element of $A$. 
	Since $k^- \leq k$, there is a component of $T[R(d_{k^-})]$ that contains two or more leaves of $A$.
	\blue{Then, by Lemma \ref{mainlemma}, the set $A$ cannot be a maxPD set, again contradicting our initial supposition.}
	Thus $A$ contains at least one leaf from each component of $T[R(d_{k^-})]$. \medskip
	
	Now suppose that $A = \{x_1,\dots,x_k\}$ contains at least one leaf from each component of $T[R(d_{k^-})]$, and at most one leaf from each component of  $T[R(d_{k^+})]$.
	By Lemma~\ref{PDupperbound}, the value $PD(\hat{T}_k) + kd_{k^-}$ is an upper bound for the PD score of size-$k$ sets.
	We show that $PD(A)$ achieves this bound.
	
	Notice that the components of $T[R(d_{k^-})]$ match those of $T^\prime_k[R(d_{k^-})]$, and the components of $T[R(d_{k^+})]$ match those of $T^\prime_k[R(d_{k^+})]$, in terms of their constituent leaves.
	For each $x_i \in A$ there exists a vertex $v_i \in V^\prime$ for which $d(v_i,x_i) = d_{k^-}$.
	Since each component of $T^\prime_k[R(d_{k^+})]$ contains at most one leaf, the paths from $v_i$ to $x_i$, and from $v_j$ to $x_j$ contain no common edges, for any distinct $1 \leq i,j \leq k$.  
	So in total, the collection of all paths $v_i$ to $x_i$ for all $i$ contributes exactly $kd_{k^-}$ to $PD(A)$.
	Furthermore, since $A$ contains at least one leaf from each component of $T[R(d_{k^-})]$, every edge of $\hat{T}_k$ is included in the $A$-subtree of $T$.
	Thus $PD_T(A) = PD_{T^\prime_k}(A) = PD(\hat{T}_k) + kd_{k^-}$, the maximum possible value, and hence $A$ is a size-$k$ maxPD set.\qed
\end{proof}

When $k$ is a branching value for a rooted phylogenetic $X$-tree $T$, then $k^- = k = k^+$ and $T[R(d_{k^-})] =T[R(d_k)] = T[R(d_{k^+})]$.
Hence, as a direct consequence of Theorem~\ref{maxPDsufficientconditions}, we obtain the following result.  

\begin{corollary}\label{branchingcomponents}
	Let $T$ be a rooted phylogenetic $X$-tree whose edge lengths satisfy the ultrametric condition.
	Let $A \subseteq X$ with $|A| = k$. 
	If $k$ is a branching value of $T$, then $A$ is a size-$k$ maxPD set if and only if $A$ contains exactly one leaf from each component of $T[R(d_k)]$.
\end{corollary}

Theorem~\ref{maxPDsufficientconditions} can be used to count the number of size-$k$ maxPD sets for a rooted phylogenetic $X$-tree $T$ whose edge lengths are ultrametric. 
Note that this result does not require $T$ to be binary.
Let $m(T,k)$ denote the number of size-$k$ maxPD sets on $T$.
The next proposition derives $m(T,k)$ when $k$ is a branching value of $T$. 
The case when $k$ is not a branching value of $T$ is covered separately. 
We express the forest $T[R(d_k)]$ as a union of components $\kappa_i(k)$ for $i \in \{1,2, \dots, k\}$, and write $\lambda(\kappa_i(k))$ for the number of leaves in $\kappa_i(k)$.

\begin{proposition}\label{mTkatbranchingvalues}
	Let $T$ be a rooted phylogenetic $X$-tree whose edge lengths satisfy the ultrametric condition. 
	If $k$ is a branching value for $T$, then $$m(T,k) = \prod\limits_{i \in \{1, \dots, k\}} \lambda(\kappa_i(k)).$$
\end{proposition}	

\begin{proof}
	By Corollary~\ref{branchingcomponents}, each maxPD set contains exactly one leaf from component $\kappa_i(k)$, for all $i \in \{1,2, \dots, k\}$.
	There are exactly $\lambda(\kappa_i(k))$ ways to choose one leaf from component $\kappa_i(k)$.
	Since the choice in each component is independent of the choices in all other components, $m(T,k) = \prod\limits_{i \in \{1, \dots, k\}} \lambda(\kappa_i(k))$.
\qed
\end{proof}

\begin{figure}
	\centering
	\includegraphics[width=0.75\textwidth]{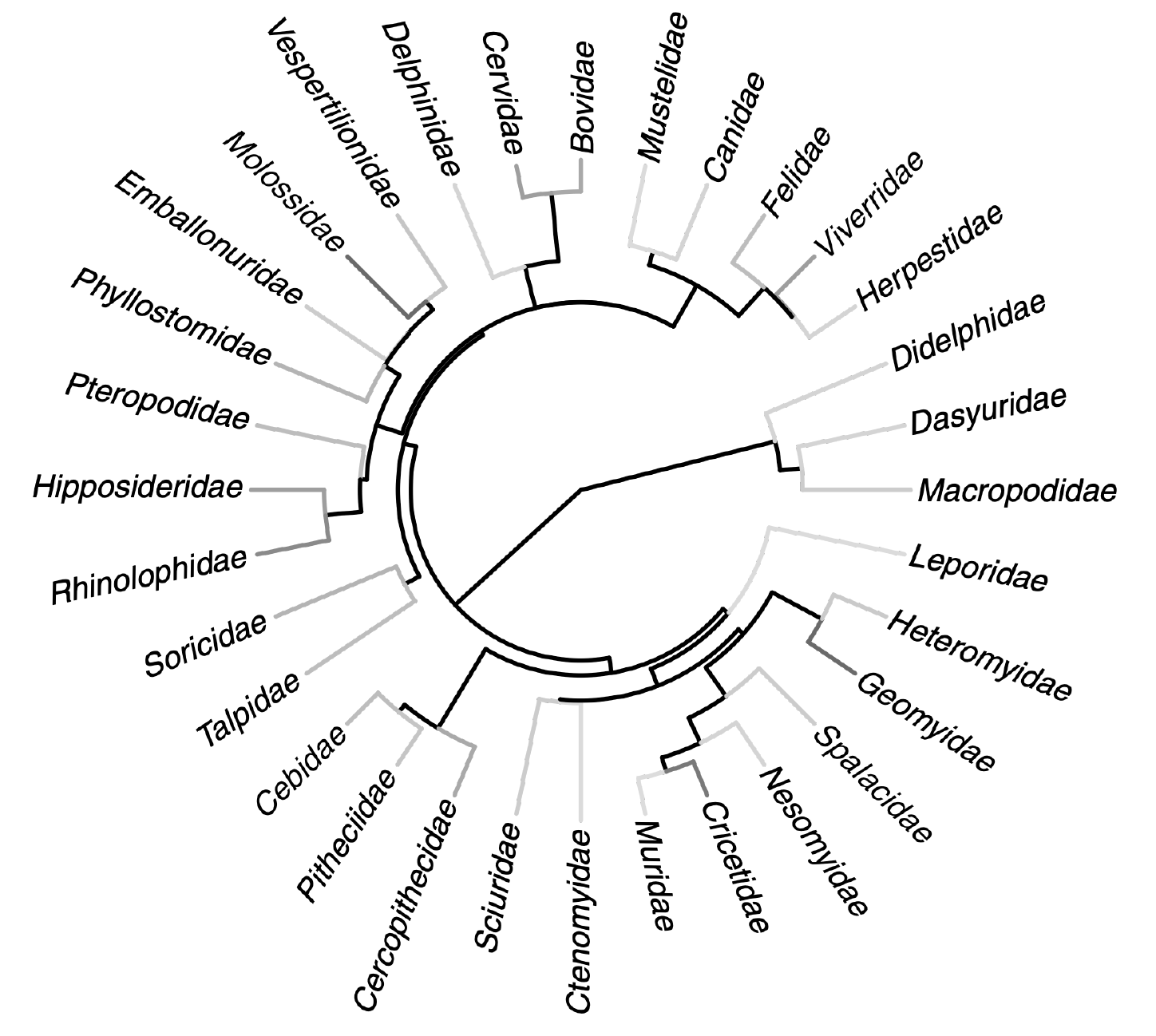}
	\caption{\blue{A phylogeny of 32 mammalian families, appearing originally as Fig.~3a in \cite{mazel2018prioritizing}.}}
	\label{mammalfamilies}
\end{figure}

\noindent\textbf{Example.}
	\cite{mazel2018prioritizing} exhibit a phylogeny of 32 mammal families \blue{(reprinted as Fig.~\ref{mammalfamilies})}. 
	Let us call this phylogeny $P$. 
	We calculate the number of size-8 and size-16 maxPD sets for $P$.
	
	The family \emph{Leporidae} appears as a single vertex component in both $P[R(d_8)]$ and $P[R(d_{16})]$. 
	For $P[R(d_8)]$, the components appearing clockwise, starting from \emph{Leporidae}, have sizes 1, 8, 3, 2, 7, 3, 5, and 3. 
	The product of these values gives a total of 15120 size-8 maxPD sets for $P$.
	Note that this represents 0.14\% of the possible sets of 8 leaves.
	For $P[R(d_{16})]$, the components appearing clockwise, starting from \emph{Leporidae}, have sizes 1, 2, 4, 1, 1, 3, 1, 1, 2, 1, 4, 3, 5, 1, 1, and 1. 
	This gives a total of 2880 size-16 maxPD sets for $P$.
	\medskip

If $k$ is not a branching value for a rooted phylogenetic $X$-tree $T$, calculating the number of size-$k$ maxPD sets is not as immediate.
In this case, we use a generating function to determine $m(T,k)$.
The following lemma is presented for a more general context.

\begin{lemma}\label{genfunction}
	Suppose that $\CC= (X_{ij}: i=1, \ldots, n_j; j=1, \ldots, r)$ is an array of disjoint sets, and let $n_{ij}$ denote the size of set $X_{ij}$.
	Let $N_\CC(k)$ be the number of sets of size $k$ that can be obtained by selecting at most one element from each set $X_{ij}$ but in such a way that at least one element is selected from $\bigcup_i X_{ij}$ for each value of $j$. 
	Then $N_\CC(k)$ is the coefficient of $x^k$ in the polynomial \begin{equation}\label{genfunction}p_{\CC, k}(x) = \prod_{j=1}^r\left(-1+\prod_{i=1}^{n_j}  (1+n_{ij}x)\right).\end{equation}
\end{lemma}

\begin{proof}
	For each integer $j \geq 1$, let
	$p_j(x)= -1+\prod_{i=1}^{n_j}  (1+n_{ij}x)$, 
	and for each $l \geq 0$, let 
	$c_{lj}$ denote the coefficient of $x^l$ in $p_j(x)$.
	Then $c_{0j}=0$ for each $j$, 
	and for $l>0$, the coefficient $c_{lj}$ is the number of ways of selecting $l$ elements from  $\bigcup_{i=1}^{n_j} X_{ij}$ in such a way that at most one 
	element is selected from the (pairwise-disjoint) sets $(X_{ij} : i=1 \ldots, n_j)$ and at least one element is selected (since $l>0$).

	Now $p_{\CC, k}(x) = \prod_{j=1}^r p_j(x)$ and so the coefficient of $x^k$ in $p_{\CC, k}(x)$ is the sum (call it $S_k$) of the terms
	$c_{l_1,1} c_{l_2,2} \cdots c_{l_r, r}$  across all  choices of $(l_1, l_2, \ldots, l_r)$ for which 
	$l_1+ l_2 + \cdots + l_r = k$, and $l_m >0$ for all $m$ (this second condition  holds  because $c_{0j}=0$ for all $j$). Since the sets
	$X_{ij}$ are pairwise-disjoint (across all choices of $i,j)$, we have $S_k = N_\CC(k)$ as required.
\qed
\end{proof}

Let $T$ be a rooted phylogenetic $X$-tree, and let $k$ be a positive integer such that $k \leq |X|$.
We write $p_{T, k}(x)$ \blue{instead of $p_{\CC,k}$} when $\CC$ is constructed from component-connected clusters of $T$, using the branching values $k^-$ and $k^+$.
If $k$ is a branching value of $T$, we have $n_j =1$ for all $j$, and the result coincides with Proposition~\ref{mTkatbranchingvalues}.
In the general case, Lemma~\ref{genfunction} gives a polynomial-time algorithm to compute $m(T,k)$.

\begin{theorem}\label{mTkpolynomialtime}
	Let $T$ be a rooted phylogenetic $X$-tree whose edge lengths satisfy the ultrametric condition.
	Let $|X| = n$, and let $k \leq n$.
	The components of $T[R(d_{k^-})]$ and $T[R(d_{k^+})]$ can be determined in time $O(n^2)$.
	The value $m(T,k)$ can be computed in time $O(n^3)$. 
\end{theorem}

\begin{proof}
There are at most $n$ different branching values for $T$ (one for every non-leaf vertex, and the value $n$) from which to select the appropriate $k^-$ and $k^+$ values. 
Determining the components of a forest can be achieved in $O(n^2)$ time.
Once the components have been determined,  the polynomial $p_{T, k}(x)$ is calculated, and the coefficient of $x^k$ is extracted.
With a na\"ive approach of sequentially multiplying factors, this can be completed in time $O(n^3)$.
\qed	
\end{proof}	
 
The following example highlights a nice property of the generating function $p_{T,k}(x)$ for a rooted phylogenetic tree $T$. 
Calculating the number of size-$k$ maxPD sets for some non-branching value $k$ also gives the number of size-$m$ maxPD sets for every positive integer $m$ in the interval $[k^-,k^+]$.\\ 

\noindent\textbf{Example.} 
	Consider the rooted phylogenetic tree $T_2$ in Fig.~\ref{fig3}.
	Firstly, 4 is a branching value for $T_2$. 
	Thus the number of its size-4 maxPD sets is the product of the number of leaves in each of the four components of $T_2[R(d_4)]$.
	That is, $m(T_2,4) = 3 \cdot 1 \cdot 4 \cdot 3 = 36$.
	
	However, 5 is not a branching value for $T_2$, so we use the generating function approach to find $m(T_2,5)$.
	\blue{First note that the greatest branching value less than 5 is 4 and the least branching value greater than 5 is 7.
	The forests $T_2[R(d_4)]$ and $T_2[R(d_7)]$ are shown in Fig.~\ref{fig3}.	
	We then construct an array of disjoint sets $\CC(T_2)$, with a view to using Eqn. \ref{genfunction}.
		Each entry of $\CC(T_2)$ consists of a set of leaves contained in one component of $T_2[R(d_7)]$.
		That is, the entries of $\CC(T_2)$ are  $\{x_1,x_2\}, \{x_3\}, \{x_4\}, \{x_5,x_6\}, \{x_7,x_8\}, \{x_9,x_{10}\}, \{x_{11}\}$.}
		
	\blue{Lastly, we arrange the entries of $\CC(T_2)$ so that each column contains precisely those leaves that share a component of $T_2[R(d_4)]$.
		Thus we have 4 columns in $\CC(T_2)$, and set $r=4$ in Eqn. \ref{genfunction}.
	 	As $T_2$ is binary, there are at most two components of $T_2[R(d_7)]$ contained in any component of $T_2[R(d_4)]$.  
	 	We use an empty set as a placeholder, if required, to ensure that $\CC(T_2)$ is a rectangular array. 
	 	As such, we are able to set $n_j = 2$ for all $j$ in Eqn. \ref{genfunction}.
	 	The completed array is
		$$\CC(T_2) = \begin{bmatrix}
		\{x_1,x_2\} & \{x_4\} & \{x_5,x_6\} & \{x_9,x_{10}\} \\
		\{x_3\} & \emptyset & \{x_7,x_8\} & \{x_{11}\} 
		\end{bmatrix}
		.$$ }
	
	The generating function for $T_2$, when $k=5$, is calculated below.
	
	\begin{align*}
	p_{T_2,5}(x) &= \prod_{j=1}^4\left(-1+\prod_{i=1}^{2}  (1+n_{ij}x)\right)\\ 
	&= [-1 + (1+2x)(1+x)]^2[-1 + (1+x)(1)][-1 + (1+2x)^2]\\
	&= x^4(2x +3)^2(4x+4)\\
	&= 16x^7 + 64x^6 + 84x^5 + 36x^4
	\end{align*}
	
	\noindent Hence $T_2$ has 84 maxPD sets of size 5.
	We have also determined that $T_2$ has 64 size-6 maxPD sets, 16 of size 7, and confirmed that there are 36 maxPD sets of size 4.

\subsection{Bounding $m(T,k)$, and its value for a certain family of trees}

The shape of a rooted phylogenetic tree impacts the components at each \blue{branching distance}, and hence the number of maxPD sets which exist.
In this section we restrict ourselves to rooted binary phylogenetic trees with the ultrametric constraint on edge lengths.
By `shape',  we refer to both the particular branching structure of a tree and the relative distances of the vertices from its leaves \citep[see][Ch. 3]{ste16}.
Here, we begin to address the question of which tree shapes and values of $k$ give the most size-$k$ maxPD sets across a fixed number of leaves.
First we consider the lower and upper bounds for the number of size-$k$ maxPD sets when $k$ is a branching value.

\begin{proposition}\label{branchingvaluebounds}
	Let $T$ be a rooted binary phylogenetic $X$-tree whose edge lengths satisfy the ultrametric condition, and let $|X| = n$. 
	If $k$ is a branching value for $T$, then
$$n - k + 1 \leq m(T,k) \leq \left(\frac{n}{k}\right)^k.$$
Moreover, these bounds are sharp.
\end{proposition}

\begin{proof}
	If $k$ is a branching value for $T$, then, by Proposition~\ref{mTkatbranchingvalues}, the value $m(T, k)$ is the product of the number of leaves in the $k$ components.
	Let $S$ be the multiset $\{\lambda(\kappa_i(k)) : 1 \leq i \leq k\}$, that is, the multiset containing the number of leaves of each component.
	
	To find the lower bound for $m(T,k)$, note that	if $a$ and $b$ are integers with $1 < a \leq b$, then $(a-1)(b+1) < ab$.
	Hence the product of elements of the multiset $(S - \{a,b\}) \cup \{a-1,b+1\}$ will be less than the product of elements of $S$.
	This exchange of elements can continue until only one element is greater than 1. 
	Thus the minimum product of $k$ positive integers which sum to $n$ is $n-k+1$, achieved with one value of $n-k+1$ and $k-1$ values of 1. 
 	This bound is achieved by rooted caterpillar trees (i.e. rooted phylogenetic $X$-trees with exactly one cherry).
	
	On the other hand, the maximum such product is bounded above by $(\frac{n}{k})^k$. 
	This follows from the fact that the arithmetic mean, $AM(S)$, of a multiset of positive integers $S$ is greater than or equal to the geometric mean, $GM(S)$, of the same multiset.
	Thus
	 $$m(T, k) = \prod_{s \in S} s = (GM(S))^k \leq (AM(S))^k = \left(\frac{n}{k}\right)^k.$$
	This maximum is obtained when $k$ is a divisor of $n$ and all components contain $\frac{n}{k}$ leaves.
	\qed
\end{proof}

Let $T$ be a rooted binary phylogenetic $X$-tree. 
\blue{If $k$ is not a branching value, it is possible that $m(T,k)$ exceeds the upper bound given in Proposition \ref{branchingvaluebounds}.
	For example, the tree $T_2$ from Fig.~\ref{fig3} has $n=11$, and $m(T_2,5) = 84 \geq \left(\tfrac{11}{5}\right)^5 \approx 51.5$.}

\blue{We have seen above that if $T$ is a rooted caterpillar tree, then $m(T,k)$ is as small as possible. 
	The highly asymmetric structure of caterpillar trees restricts the possible maxPD sets they contain. 
	In contrast, we now consider $m(T,k)$ values across the family of fully symmetric rooted trees (with constant edge lengths) and include cases when $k$ is not a branching value.}

We say $T$ is a \emph{perfect unit-length} tree if the edge lengths of $T$ satisfy the ultrametric condition, and all edges of $T$ have length 1.
Perfect unit-length trees have $2^\alpha$ leaves, where $\alpha \in \NN$ is the \emph{height} of the tree (the number of edges between the root and any leaf).

\begin{proposition}
	Let $T$ be a perfect unit-length tree of height $\alpha \in \NN$, and let $n$ denote the number of leaves of $T$.
	Let $k$ be a positive integer such that $k \leq n$, and let $\beta$ be the unique non-negative integer such that $2^{\beta-1} < k \leq 2^\beta$.
	Then 
	\begin{equation}
	\label{minty}
	m(T,k) = {2^{\beta-1} \choose k - 2^{\beta-1}} \cdot 2^{2^\beta + (\alpha-\beta - 1)k}.
	\end{equation}
	The values of $k$ that maximise $m(T,k)$ are $k= \lfloor \frac{2n}{3} \rfloor$ for all $n$ and, additionally, $k = \lfloor \frac{2n}{3} \rfloor + 1$ when $n \equiv 1 \pmod 3$.
\end{proposition}

\begin{proof}
	Firstly, if $k$ is a branching value, then $k=2^\beta$ and each component has size $2^{\alpha-\beta}$.
	Therefore, by Proposition~\ref{mTkatbranchingvalues}, $m(T,k) = (2^{\alpha-\beta})^k$, which coincides with Eqn. (\ref{minty}).
	
	Furthermore, if $k$ is not a branching value, we have $k^- = 2^{\beta - 1}$ and $k^+ = 2^\beta$. 
	Then by Lemma~\ref{genfunction}, $m(T,k)$ is the coefficient of $x^k$ in the polynomial 
	 $$p_{T,k}(x) = (-1 + (1 + 2^{\alpha -\beta}x)^2)^{2^{\beta-1}} = (2^{2(\alpha -\beta)}x^2 + 2^{\alpha -\beta + 1}x)^{2^{\beta-1}}.$$ 
	Taking the binomial expansion of the last expression we determine that ${2^{\beta-1} \choose k - 2^{\beta-1}} \cdot 2^{2^\beta + (\alpha-\beta - 1)k}$ is the coefficient of $x^k$ in $p_{T,k}(x)$.
	This establishes Eqn. (\ref{minty}).
	
	To find the value of $k$ which maximises $m(T,k)$, we first show that $m(T,k) \leq m(T,n-k)$ when $k \leq \frac{n}{2}$.
	Let $A$ be a size-$k$ maxPD set for some $k \leq \frac{n}{2}$.
	By Theorem~\ref{maxPDsufficientconditions}, $A$ contains at most one leaf from each cherry of $T$.
	Then $X-A$ contains at least one leaf from every cherry (which are the components of $T[R(d_{(n-k)^-})]$), and at most one leaf from each component of $T[R(d_{(n-k)^+})]$, as these are all single-leaf components.
	This implies $X - A$ is a size-$(n-k)$ maxPD set by Theorem~\ref{maxPDsufficientconditions}, and thus there are at least as many size-$(n-k)$ maxPD sets as size-$k$ ones.
	Hence the value of $k$ that maximises $m(T,k)$ will be greater than or equal to $\frac{n}{2}$.
	
	In the case when $k \geq \frac{n}{2}$, since $2^\alpha = 2^\beta = n$, the expression in  Eqn.~(\ref{minty}) simplifies to $m(T,k) = {{\frac{n}{2}} \choose {k - \frac{n}{2}}} \cdot 2^{n-k}$. 
	Computing the ratio $\frac{m(T, k+1)}{m(T, k)}$, we have
	
	$$\frac{m(T,k+1)}{m(T,k)} = \frac{{{\frac{n}{2}} \choose {k - \frac{n}{2} +1}} \cdot 2^{n-k-1}}	  {{{\frac{n}{2}} \choose {k - \frac{n}{2}}} \cdot 2^{n-k}} = \frac{n-k}{2k-n +2}.$$
	
	This ratio is monotonically decreasing as $k$ increases, and equals 1 when $3k = 2n-2$.
	Our maximal value of $m(T, k)$ will be found at the smallest $k \geq \frac{n}{2}$ for which $\frac{m(T, k+1)}{m(T, k)} \leq 1$, namely when $k= \lfloor \frac{2n}{3} \rfloor$. 
	Note that when $n \equiv 1 \pmod 3$ and $k = \lfloor \frac{2n}{3} \rfloor$, we have $\frac{m(T, k+1)}{m(T,k)} = 1$, so we get an equal number of maxPD sets for the two consecutive values $k$ and $k+1$.  
	\qed
\end{proof}

\begin{table}
	\begin{center}	
		\begin{tabular}{| c | c | c |}
			\hline
			$n$ & $k$ & $m(T,k)$\\
			\hline
			4 & 1,2,3 & 4\\
			8 & 3,5 & 32\\
			16 & 10,11 & 1,792\\
			32 & 21 & 8,945,664\\
			64  & 42,43 & $\sim 2.7 \times 10^{14}$\\
			\hline
			
		\end{tabular}
	\end{center}
	\caption{Number of size-$k$ maxPD sets for a perfect unit-length tree $T$ with $n$ leaves}
\end{table}

Table 1 shows the growth of $m(T, k)$ as $n$ increases.
We note that for $n = 16$, the perfect unit-length tree does not provide the largest value of $m(T, k)$.
Figure~\ref{sixteenmax} shows a rooted binary phylogenetic $X$-tree $T_3$ on 16 leaves which contains 1809 size-8 maxPD sets (thus having 17 more maxPD sets than the perfect unit-length tree on 16 leaves can achieve for its optimal value of $k=10$ or $k=11$). For $T_3$ we have $p_{T_3,8}(x) = (x^2+2x)^2(2x^2+3x)^4$. 

\begin{figure}
	\centering
	\begin{tikzpicture}[scale=0.7]
	
	\begin{scope}[every node/.style=vertex]
	\node at (2,5) (t) [draw=none,fill=none]{$T_3$};
	\node at (7.25,5.375) (rho) [label=above:$\rho$]{};
	\node at (0,0) (x1) [label=below:$x_1$]{};
	\node at (1,0) (x2) [label=below:$x_2$]{};
	\node at (2,0) (x3) [label=below:$x_3$]{};
	\node at (3,0) (x4) [label=below:$x_4$]{};
	\node at (4,0) (x5) [label=below:$x_5$]{};
	\node at (5,0) (x6) [label=below:$x_6$]{};
	\node at (6,0) (x7) [label=below:$x_7$]{};
	\node at (7,0) (x8) [label=below:$x_8$]{};
	\node at (8,0) (x9) [label=below:$x_9$]{};
	\node at (9,0) (x10) [label=below:$x_{10}$]{};
	\node at (10,0) (x11) [label=below:$x_{11}$]{};
	\node at (11,0) (x12) [label=below:$x_{12}$]{};
	\node at (12,0) (x13) [label=below:$x_{13}$]{};
	\node at (13,0) (x14) [label=below:$x_{14}$]{};
	\node at (14,0) (x15) [label=below:$x_{15}$]{};
	\node at (15,0) (x16) [label=below:$x_{16}$]{};
	\node at (5.75,4.625) (a) {};
	\node at (4.25,3.875) (b) {};
	\node at (2.75,3.125) (c) {};
	\node at (1.5,2.5) (d) {};
	\node at (0.5,2) (e) {};
	\node at (2.5,2) (f) {};
	\node at (5,2) (g) {};
	\node at (8,2) (h) {};
	\node at (11,2) (i) {};
	\node at (14,2) (j) {};
	\node at (5.5,1) (k) {};
	\node at (8.5,1) (l) {};
	\node at (11.5,1) (m) {};
	\node at (14.5,1) (n) {};
	\end{scope}
	
	\path[draw,thick]
	(rho) edge (a)
	(rho) edge (j)
	(a) edge (b)
	(a) edge (i)
	(b) edge (c)
	(b) edge (h)
	(c) edge (d)
	(c) edge (g)
	(d) edge (e)
	(d) edge (f)
	(e) edge (x1)
	(e) edge (x2)
	(f) edge (x3)
	(f) edge (x4)
	(g) edge (x5)
	(g) edge (k)
	(h) edge (x8)
	(h) edge (l)
	(i) edge (x11)
	(i) edge (m)
	(j) edge (x14)
	(j) edge (n)
	(k) edge (x6)
	(k) edge (x7)
	(l) edge (x9)
	(l) edge (x10)
	(m) edge (x12)
	(m) edge (x13)
	(n) edge (x15)
	(n) edge (x16)
	;
	\end{tikzpicture}
	\caption{A rooted binary phylogenetic tree with more maxPD sets (for its optimal value of $k=8$) than the perfect unit-length tree with the same number of leaves (for its optimal value of $k = 10,11$)}.
	\label{sixteenmax}
\end{figure}

\section{Finding a maxPD set that maximises a linear function on the leaves}
\label{sec-linearfunction}

Section~\ref{sec_maxPD} presented methods for determining the number of size-$k$ maxPD sets for a given rooted phylogenetic tree. 
These methods confirmed the observations in the literature that, in general, maxPD sets are far from unique.
This provides scope for evaluating the collection of maxPD sets against other strategic considerations.
In developing strategies for conservation planning, PD is often seen as one measure to be used in conjunction with others (for examples of this, see
\cite{cadotte2018difficult, isaac2007mammals, kling2019facets}).
For instance, we may wish to incorporate benefit-cost ratios of focussed conservation spending, or employ IUCN categorisations into the analysis. 
This section provides an algorithm to optimise a further measure across the collection of maxPD sets. 

Here, we frame the further measure in terms of a real-valued linear function on the leaves.
Each leaf is assigned a function value, and the \emph{linear function score} of a set of leaves is the \blue{weighted} sum of the function values of the constituent leaves.
We seek a size-$k$ set which has as large a linear function score as possible among the size-$k$ maxPD sets.
\blue{By suitably modifying the linear function, the problem can be rephrased as maximising the unweighted sum across maxPD sets. Thus,} for a function $\phi:X \rightarrow \RR$ we want to determine $$\max\left\{\sum_{x \in A}\phi(x) : A \textrm{~is~a~size-}k \textrm{~maxPD~set}\right\}.$$

We note that it is not always possible to achieve this result by simply adding the function score of each leaf to the length of its incident pendant edge, and then finding a size-$k$ maxPD set of the resulting rooted phylogenetic tree.
We provide a counterexample using the tree $T_1$ from Fig.~2.
Consider the function
$$f(x) = 
\begin{cases}
1, & \mbox{ if } x \in \{x_1, x_2, x_3, x_4\};\\
100,  & \mbox{ if } x \in \{x_5, x_6, x_7\}.
\end{cases} 
$$
Adding the function values to appropriate pendant edges, results in a tree with a unique size-3 maxPD set $\{x_5, x_6, x_7\}$. 
However this set is not a maxPD set of the original tree $T_1$. 

For a rooted phylogenetic tree $T$, \textsc{MaximiseLinearSum} selects a set $A$ consisting of $k$ leaves of $T$ in the following manner.
Initially, it determines the components of $T[R(d_{k^-})]$ (`tall' components) and those of $T[R(d_{k^+})]$ (`short' components). 
For every short component, it keeps (in the set of `potential' leaves $P$) one leaf $x$ such that $\phi(x)$ is maximal for that short component.
It then discards all other leaves from further consideration.
In every tall component, it adds one leaf $x$ to $A$ from the leaves retained in $P$ such that $\phi(x)$ is maximal for that tall component.
Finally, from the remaining $k^+ - k^-$ leaves under consideration, it chooses $k - k^-$ with the largest $\phi$ values.
In presenting the algorithm, we make use of the following notation.
For a pendant subtree $C$ of $T$, write $X_C$ for the set of leaves in $C$. 
For $S \subseteq X$, let $\phi(S) = \{\phi(x): x \in S\}$.

\IncMargin{1em}
\begin{algorithm}
	\caption{\textsc{MaximiseLinearSum}}\label{alglinearfunction}
	\SetKwInOut{Input}{Input}\SetKwInOut{Output}{Output}
	\normalsize
	
	\KwIn{a rooted phylogenetic $X$-tree $T$ whose edge lengths satisfy the ultrametric condition,\\ \hspace{11mm} a positive integer $k \leq |X|$,\\ \hspace{11mm} $\phi: X \rightarrow \mathbb{R}$}
	\KwOut{a size-$k$ maxPD subset $A \subseteq X$, with the largest linear function score among all maxPD sets}
	\BlankLine
	determine $T[R(d_{k^-})]$ and $T[R(d_{k^+})]$\;
	$P \leftarrow \emptyset$ \tcc*{Potential leaves to include}
	$A \leftarrow \emptyset$ \tcc*{Output set}
	\ForEach{component $C$ in $T[R(d_{k^+})]$}{
		choose one leaf $m$ from the set $\{x \in X : \phi(x) = \max\phi(X_C)\}$\;
		$P \leftarrow P \cup m$
	}
	\ForEach{component $C$ in $T[R(d_{k^-})]$}{
		choose one leaf $m$ from the set $\{x \in P : \phi(x) = \max\phi(X_C \cap P)\}$\;
		$A \leftarrow A \cup m$\;
		$P \leftarrow P - m$
	}	
	for each of the $k - k^-$ largest elements $\phi(x)$ of $\phi(P)$ add $x$ to $A$\;
	\KwRet{$A$}		
	
\end{algorithm}
\DecMargin{1em}

\begin{proposition}
	Let $T$ be a rooted phylogenetic $X$-tree whose edge lengths satisfy the ultrametric condition.
	The \textsc{MaximiseLinearSum} algorithm outputs a maxPD set of $T$.
\end{proposition}

\begin{proof}
	The for-loop from Lines 4 to 7 ensures that $A$ cannot contain more than one leaf from any short component. 
	The for-loop from Lines 10 to 13 ensures that $A$ contains at least one leaf from every large component. 
	Since Line 15 ensures that $|A| =k$, it follows by Theorem~\ref{maxPDsufficientconditions}, that $A$ is a maxPD set of $T$. \qed
\end{proof}

\begin{proposition}
	Let $T$ be a rooted phylogenetic $X$-tree whose edge lengths satisfy the ultrametric condition, and let $\phi : X \rightarrow \RR$ be a function on the leaves of $T$.
	Let $k$ be a positive integer such that $k \leq |X|$.
	Then \textsc{MaximiseLinearSum} applied to $T$, $\phi$, and $k$ correctly outputs a size-$k$ maxPD set with the largest function score among all maxPD sets.
\end{proposition}

\blue{We give a proof of this result shortly, but first give a short description of our approach. The algorithm \textsc{MaximiseLinearSum} was designed to construct a set containing the largest possible values of $\phi$ while obeying the constraints imposed by Theorem \ref{maxPDsufficientconditions} to ensure the selection of a maxPD set. Suppose that $A$ is a size-$k$ maxPD set of $T$. The proof considers two possible cases when $A$ is not a valid output of the algorithm, and exhibits a size-$k$ maxPD set with a greater linear function score in each. Finally, outside of these two cases we prove that the linear function score of $A$ must be at least as large as that of any other size-$k$ maxPD set.}

\begin{proof} 
	Suppose that $A$ is a size-$k$ maxPD set of $T$. 
	For the result to hold, either $A$ is a valid output of \textsc{MaximiseLinearSum} or there is a size-$k$ maxPD set $B$, distinct from $A$, such that $\sum_{x \in A} \phi(x) < \sum_{x \in B} \phi(x)$.
	One of the following three conditions holds:
	\begin{enumerate}
		\item There is a component of $T[R(d_{k^+})]$ (a short component) which contains leaf $a \in A$ and leaf $b \in B$, where $\phi(a) < \phi(b)$.
		In this case, $a$ would not be selected in Line 5 of \textsc{MaximiseLinearSum}, meaning $A$ cannot be a valid output of this algorithm.
		However, the set $A^\prime = (A - a) \cup b$ is a size-$k$ maxPD set with $\sum_{x \in A} \phi(x) < \sum_{x \in A^\prime} \phi(x)$.
		\item Condition 1 fails, and there is a component of $T[R(d_{k^-})]$ (a tall component) which contains leaves $\{a_1,a_2,\dots,a_s\} \subseteq A$, and $\{b_1,b_2,\dots,b_t\} \subseteq B$ where, for some $j \in \{1,2,\dots,t\}$, the inequality $\phi(b_j)>\phi(a_i)$ holds across all $i~\in~\{1,2,\dots,s\}$.
		In particular, the strictness of this inequality means that  $b_j \notin A$.
		In this case, the leaf $b_j$ would always be selected by Line 9 of \textsc{MaximiseLinearSum}, precluding $A$ from being a valid output of this algorithm.
		Moreover, since Condition 1 fails to hold, no element of $\{a_1,a_2,\dots,a_s\}$ shares a short component with $b_j$.
		Thus $A^\prime = (A - a_1) \cup b_j$ is a size-$k$ maxPD set with $\sum_{x \in A} \phi(x) < \sum_{x \in A^\prime} \phi(x)$.
		\item Conditions 1 and 2 fail. 
		Thus, in every component of $T[R(d_{k^-})]$, the set $A$ contains a leaf that has the maximal $\phi$ value for that component.
		Assume that $k$ is a branching value of $T$. 
		Then $A$ is a valid output of \textsc{MaximiseLinearSum}, as the choice applied in Line 9 can be the one element of $A$ from within each component. 
		Additionally, since elements of $A$ have the maximal $\phi$ value in each component, $\sum_{x \in A} \phi(x) \geq \sum_{x \in B} \phi(x)$ for any size-$k$ maxPD set $B$.
		Thus the proposition holds when $k$ is a branching value of $T$.
		
		Now assume that $k$ is not a branching value.
		Let $\bar A \subseteq A$ consist of $k^-$ elements of $A$ which have the maximal $\phi$ value in their tall component, one from each tall component.
		We construct the set $\bar B \subseteq B$ to include (i)~elements of $B$ that share a short component with some leaf in $\bar A$, and (ii)~from tall components where no leaf in $B$ satisfies Condition (i), one element of $B$ in each such tall component with the largest $\phi$ value. 
		The set $\bar B$ contains exactly one leaf from each tall component. 
		For $a \in \bar A$ and $b \in \bar B$ in the same tall component, $\phi(a) \geq \phi(b)$.
		Thus \begin{equation}\label{barAbarB} \sum_{x \in \bar A} \phi(x) \geq \sum_{x \in \bar B} \phi(x). \end{equation}
	
		Let $P$ be the set of `potential leaves' as used in Algorithm \ref{alglinearfunction}. 
		Then by our construction of $\bar{B}$, we have $B - \bar B \subseteq P - \bar A$.
		The set $A$ is a valid output of \textsc{MaximiseLinearSum} if and only if the elements of $A - \bar A$ have the $k-k^-$ largest $\phi$ values among elements of $P-\bar A$.
		The latter condition is equivalent to $\sum_{x \in A - \bar A} \phi(x) \geq \sum_{x \in B - \bar B} \phi(x)$, that is $\sum_{x \in A} \phi(x) \geq \sum_{x \in B} \phi(x)$.
	\end{enumerate}

Hence, under all three conditions, either $A$ is a valid output of \textsc{MaximiseLinearSum} or $\sum_{x \in A} \phi(x) < \sum_{x \in B} \phi(x)$ for some size-$k$ maxPD set $B$ of $T$, as required.
\qed
\end{proof}

\begin{proposition}
	Let $T$ be a rooted phylogenetic $X$-tree whose edge lengths satisfy the ultrametric condition, and let $|X| = n$.
	Then \textsc{MaximiseLinearSum} runs in time $O(n^2)$.
\end{proposition}

\begin{proof}
	By Theorem~\ref{mTkpolynomialtime}, Line 1 can be completed in $O(n^2)$. 
	We show that this subroutine dominates the time taken for \textsc{MaximiseLinearSum} to run.
	
	Determining which vertices are in each component can be achieved by a depth-first search in linear time. 
	Both for-loops are completed in $O(n^2)$ time, as there are at most $n$ components and a component contains at most $n$ leaves. 
	Sorting a set and returning the $k - k^-$ largest values can be achieved in $O(n \log n)$ time.
	Hence, \textsc{MaximiseLinearSum} runs in the same order of time as determining the components of $T[R(d_{k^-})]$ and $T[R(d_{k^+})]$. \qed
\end{proof}

The algorithm \textsc{MaximiseLinearSum} makes use of the component constraints on maxPD sets to solve this problem for rooted phylogenetic $X$-trees whose edge lengths satisfy the ultrametric condition. 
\blue{For phylogenetic trees whose edge lengths do not satisfy the ultrametric condition, the determination of appropriate connected components requires a further algorithm \citep[in preparation]{mansonthesis}.}

\blue{We note that an alternative approach to solving this more general problem comes from the area of lexicographic multi-objective linear programming \citep[see Section 2]{cococcioni2018lexicographic}. 
The optimisation can be phrased as a max-flow min-cost problem, in a similar manner to that used in \citep{bordewich2009optimizing}.
However this approach relies on first scaling every length of the phylogenetic tree by a suitably large number. 
Determining an appropriate value for the scaling factor can prove difficult unless the edge lengths are restricted to take only rational values.
For some trees with real-valued edge lengths this step requires a pairwise comparison of the PD scores across all sets of $k$ leaves \citep[in preparation]{mansonthesis}.}

\section{Maximum possible loss of PD in a tree if $k$ species become extinct (`minPD')}
\label{sec-minPD}

In Section~\ref{sec_maxPD}, we were interested in finding sets of $k$ species which contained as much diversity as possible. 
However, it is also worth considering the dual problem: determining how must PD could be lost if $k$ extant species were to become extinct (i.e. a `worst case scenario' in biodiversity conservation in the face of widespread extinction pressure). More precisely, we consider the problem of determining the  \emph{maximum}  possible PD loss if a given number species were to become extinct. 

Formally, let $T = (V,E)$ be a rooted phylogenetic $X$-tree and let the function $\ell: E(T) \rightarrow \mathbb{R}^{>0}$ assign a positive real-valued length $\ell(e)$ to each edge $e \in E(T)$. Suppose that each species in a subset $Y$ of the leaf set $X$ of $T$ is removed from the tree.
The resulting loss of PD, which we denote here as $\Delta_{(T, \ell)}(Y)$ is given by
$$\Delta_{(T, \ell)}(Y)= PD_{(T, \ell)}(X) - PD_{(T, \ell)}(X - Y).$$
This is equivalent to  the concept of `exclusive molecular phylodiversity' as described in \cite{lew05}.\footnote{The function $\Delta_{(T, \ell)}$ is a supermodular (and decreasing) function on the  lattice of subsets of $X$, since $PD$ is a submodular (and increasing) function on this same lattice \citep{ste16}.}

Notice that finding a subset $Y$ of $X$ of size $k'$ to maximise $\Delta_{(T, \ell)}(Y)$ is equivalent to finding a subset $W$ ($=X - Y)$ of $X$ of size $k = |X|-k'$ to minimise $PD_{(T, \ell)}(W)$.   Unlike the max-PD question, this minimisation question is not solved by the greedy algorithm \citep{mou07}. 
\blue{However, as discussed in Section 6 of \citep{spillner2008computing}, minimal PD scores can be found using dynamic programming. In particular, \citep[Section 3.1]{blum1994minimum} describe an algorithm for an equivalent problem (referred to as the \emph{i-tree problem}). Here we present a detailed description of this algorithm using the terminology of phylogenetic trees.}

We call a set of $k$ leaves which has the smallest PD score across all sets of size $k$, a \emph{size-$k$ minPD set}.  In this section, we present a polynomial-time dynamic programming approach to finding minPD scores. For simplicity, we initially restrict our attention to rooted binary phylogenetic trees; however, we show that the same idea extends to rooted phylogenetic trees for which each vertex has bounded out-degree. Note that in this section, we do not require the branch lengths to satisfy the ultrametric condition. 

Given a rooted phylogenetic $X$-tree $T$, and an integer $0 \leq k \leq |X|$, let $\varphi_T(k)$ be the minimum  PD score across all size-$k$ subsets of $X$.
When $k > |X|$,  $\varphi_T(k)$ is undefined, and when $k = 0$, we set $\varphi_T(0) = 0$.
For the case when $T$ is a single vertex, we define $\varphi_T(k) = 0$.
Proposition~\ref{binaryminPD} gives the dynamic programming equation when $T$ is binary.

\begin{proposition} \label{binaryminPD}
	Let $T$ be a rooted binary phylogenetic $X$-tree and let $e_1$ and $e_2$ be the two edges of $T$ incident with the root.
	Let $e_1$ have length $\ell_1$ and $e_2$ have length $\ell_2$. 
	Finally, let $T_1$ and $T_2$ denote the (maximal) pendant subtrees formed by the deletion of $e_1$ and $e_2$ respectively. 
	
	\noindent For all $k \in \{1,2, \ldots, |X|\}$,
	\begin{equation}\label{minPDdynamiccalc}\varphi_T(k) = \min_{\substack{k_1, k_2 \geq 0,\\ k_1 + k_2 = k}} \{\varphi_{T_1}(k_1) +\varphi_{T_2}(k_2) + \ell_1\cdot \II_{k_1>0} + \ell_2\cdot \II_{k_2 >0}\},\end{equation}
	where $\II_{k_j>0}$ takes the value 1 if $k_j>0$; otherwise,  $\II_{k_j>0} = 0$.
\end{proposition}

\begin{proof}
	We proceed by induction on the number of vertices in $T$.
	For the base case, take the tree $T$ consisting of a single vertex. 
	Since $T$ has no edges, it has a PD score of 0, which corresponds to $\varphi_T(k)$ for all $k \geq 0$ by definition. 
	
	Suppose that Eqn.~(\ref{minPDdynamiccalc}) fails to give the minimum PD score for some rooted binary phylogenetic $X$-tree $T$. 
	We write $\varphi_i$ as shorthand for  $\varphi_{T_i}$ for $i=1,2$.
	Furthermore, suppose that $\varphi_i(k^\prime)$ equals the size-$k^\prime$ minPD score in $T_i$ for all $k^\prime \leq k$ and $i \in \{1,2\}$.
	Since Eqn.~\ref{minPDdynamiccalc} fails, there must be a set of $k$ leaves of $T$ which has a lower PD score than any value in the set $$\{\varphi_1(k_1) +\varphi_2(k_2) + \ell_1\cdot \II_{k_1>0} + \ell_2\cdot \II_{k_2 >0} : k_1, k_2 \geq 0, k_1 + k_2 = k\}.$$ 
	\noindent Let $A$ be such a set of $k$ leaves of $T$, with $k_1$ leaves in $T_1$ and $k_2$ leaves in $T_2$. 
	
	If $k_2 = 0$, then $PD_T(A) = \ell_1 + PD_{T_1}(A) \geq \ell_1 + \varphi_i(k_1)$ by the inductive assumption. 
	Thus the PD score of $A$ is not lower than the calculated minimum; hence, $k_2 \neq 0$.
	Similarly, $k_1 \neq 0$.
	Consequently, 
$$PD_T(A) = \ell_1 + \ell_2 + PD_{T_1}(A \cap T_1) + PD_{T_2}(A \cap T_2).$$
	For $A$ to have a PD score lower than $\varphi_T(k)$, we must have $$PD_{T_1}(A \cap T_1) + PD_{T_2}(A \cap T_2) < \varphi_1(k_1) +\varphi_2(k_2).$$
	This implies that either $PD_{T_1}(A \cap T_1) < \varphi_1(k_1)$ or $PD_{T_2}(A \cap T_2) < \varphi_2(k_2)$, contradicting our inductive assumption.
	Therefore, no such set $A$ exists, and $\varphi_T(k)$ calculates a minPD score of size $k$ in $T$.
	\qed
\end{proof}

We now present an algorithm which utilises Proposition~\ref{binaryminPD} to calculate a minPD score for a rooted binary phylogenetic $X$-tree $T=(V,E)$. 
For a vertex $v \in V(T)$, we use the notation $\varphi_v(k)$ in place of $\varphi_{T_v}(k)$, where $T_v$ is the pendant subtree of $T$ for which vertex $v$ has in-degree 0. Additionally, $\varphi_\rho(k) = \varphi_T(k)$.
Note that the root vertex $\rho$ of $T$ will always appear last in the ordered list $L$ defined in the algorithm.
For a positive integer $i$, let $L[i]$ denote the $i$th entry in list $L$.

\IncMargin{1em}
\begin{algorithm}[H]
	\caption{\textsc{MinPDScore}}\label{minPDalg}
	\SetKwInOut{Input}{Input}\SetKwInOut{Output}{Output}
	\normalsize
	
	\Input{a rooted binary phylogenetic $X$-tree $T=(V,E)$, with root $\rho$,\\ an integer $0 \leq k \leq |X|$.}
	\Output{a real number $\varphi_T(k)$}
	\BlankLine
	\ForEach{$x \in X$}{
		$\varphi_x(0) \leftarrow 0$\;
		$\varphi_x(1) \leftarrow 0$
	}
	$L \leftarrow$ ordered list of vertices in $V(T) - X$ such that if $u$ is a descendant of $v$, then $u$ appears before $v$\;
	$i \leftarrow 1$\;
	$j \leftarrow 0$\;
	\While{$i < |V(T) - X|$}{
		\ForEach{$0 \leq j \leq k$}{
			calculate $\varphi_{L[i]}(j)$ according to Eqn.~(\ref{minPDdynamiccalc}) in Proposition~\ref{binaryminPD}\;
		}
		$i \leftarrow i + 1$\;
	}
	\KwRet{$\varphi_\rho(k)$}		
	
\end{algorithm}
\DecMargin{1em}

\noindent The algorithm \textsc{MinPDScore} computes the minimum PD score for a rooted binary phylogenetic tree $T$ when selecting $k$ of its leaves.
This dynamic programming approach calculates the minPD score for pendant subtrees of $T$, which are then combined to calculate the minPD score for $T$ as a whole.
Additionally, by tracking the indicator function values as we go, a corresponding size-$k$ minPD set can be determined.

\begin{proposition}
	Let $T = (V,E)$ be a rooted binary phylogenetic $X$-tree, and let $0 \leq k \leq n$, where $n = |X|$. 
	The algorithm \textsc{MinPDScore} applied to $T$ and $k$ calculates the minimum PD score for a size-$k$ set of leaves of $T$ in time $O(n^4)$.
\end{proposition}

\begin{proof}
	Let $|X| = n$. 
The ordering of vertices on Line 5 can be completed in $O(|V(T) - X|+|E(T)|) = O(n)$ \citep{kahn1962topological}. 
The ``while" loop from Lines 8 to 13 has order $O(n) \cdot O(n) \cdot O(n^2) = O(n^4)$, since  $|V(T) - X| = n-1$, and $k \leq n$, and we are comparing $k+1$ values in Eqn.~(\ref{minPDdynamiccalc}).   
\qed
\end{proof}

\subsection{minPD scores for non-binary rooted phylogenetic trees}

The algorithm \textsc{MinPDScore} can be adapted for a non-binary rooted phylogenetic tree with bounded out-degree.
Specifically, Line 10 of the algorithm is adjusted, and an upper bound on the out-degree of every vertex is required to ensure that the modified algorithm runs in polynomial time.

Let $\{e_1,e_2,\dots,e_t\}$ denote the set of edges incident with the root of $T$, and let $T_i$ denote the subtree of $T$ descending from $e_i$.
Set  $\ell_i = \ell(e_i)$ for $i \in \{1,2,\dots,t\}$, and let
 $$K(k,t) = \left\{{\bf k}=(k_1,..., k_t): k_i \geq 0 {\rm~for~all~} i {\rm ~and~} \sum_{i=1}^t k_i = k\right\}.$$
Then, in place of Eqn. (\ref{minPDdynamiccalc}), we use Eqn. (\ref{nonbinminPDeqn}) which applies the same notation as Proposition~\ref{binaryminPD}. \begin{equation}\label{nonbinminPDeqn}\varphi_T(k) = \min_{{\bf k} \in K(k,t)} \left\{\sum_{i=1}^t \left( \varphi_i(k_i) + \ell_i\cdot \II_{k_i>0} \right)\right\}\end{equation}

\section{Concluding Remarks}

Phylogenetic diversity provides a formal way to quantify recent (and possible future) biodiversity loss,  resulting from the current high rate of species extinction.  For example, PD has become an integral part of the  Zoological Society of London’s `EDGE of Existence' programme for monitoring biodiversity risk \citep{isaac2007mammals}.  PD is more nuanced than simply counting species extinctions, since PD explicitly incorporates the evolutionary relationships among species, and thus provides a proxy for  measuring the richness of features that make species unique \citep{fai92, wicke2021formal}.

In this paper, we have investigated new combinatorial questions concerning PD that arise  in its application to large data-sets. In particular, we have described a precise way to count the number of maxPD sets of given size on a given tree (in the usual  ultrametric setting) and derived some bounds on the growth rate for these numbers.  We have also described further mathematical results that establish polynomial-time algorithms to (i) optimise a linear function (across the species at the tips of the tree) over all maxPD sets  and (ii) determine the greatest possible loss of PD on a tree if $k$ species were to become extinct (this last question amounts to determining minPD sets of given size). 

Our results suggest a number of questions.   In future work, we hope to characterise the tree shapes that have the largest number of maxPD sets (of any given size). 
A further question is to count the number of minPD sets in the binary ultrametric setting.  For caterpillars on $n$ leaves (and ultrametric edge lengths), the number of size-$k$ minPD sets is $1$ unless $k=1$, in which case there are $n$ min PD sets. To see this, observe that a size-$k$ minPD set in a caterpillar is the one that contains the $k$ leaves with the shortest pendant edges (removing any of these leaves to replace it with one of the $n-k$ unchosen leaves would necessarily add more to the PD score than what was lost by not counting the removed pendant edge).  A related question is to categorise the trees which have a {\em unique} minPD set. 

\section{Acknowledgements}
The authors were supported by the New Zealand Marsden Fund (MFP-UOC2005).
\blue{We thank the referees for helpful suggestions and for alerting us to some relevant previous papers.}
We also thank Arne Mooers for some helpful comments and suggestions.

\bibliographystyle{model2-names}
\bibliography{KMbibliography}

\end{document}